\documentclass[global,twocolumn]{svjour}

\usepackage{latexsym}
\usepackage{epstopdf}
\usepackage{amsmath}
 \usepackage{graphicx}
\journalname{Springer}

\begin{document}
\title{Engineering of Landau--Zener tunneling}

\author{Ghazal Tayebirad \inst{1} \and Riccardo Mannella \inst{2} \and Sandro Wimberger \inst{1,3}}
\institute{Institut f\"ur Theoretische Physik, Universit\"at Heidelberg, Philosophenweg 19, 69120 Heidelberg, Germany
  \and CNR-INFM and Dipartimento di Fisica `E. Fermi', Universit\`{a} di 
Pisa, Largo Pontecorvo 3, 56127 Pisa, Italy
  \and Center for Quantum Dynamics, Universit\"at Heidelberg, Germany}

\offprints{Ghazal Tayebirad\\
\email{g.Tayebirad@thphys.uni-Heidelberg.de}}

 \date{Received: date / Revised version: date}
\date{\today}

\maketitle

\begin{abstract}
Several ways are discussed how to control the Landau--Zener tunneling in the Wannier--Stark system. We focus on a realization of this system with interacting and noninteracting ultracold bosons. The tunneling from the ground band to the continuum is shown to depend crucially on the initial condition and system parameters and,  more interestingly, on added time-dependent disorder -- noise -- on the lattice beams.
\keywords {Quantum Tunneling -- Bose--Einstein Condensate -- Driven Systems -- Noise}
\end{abstract}

\section{Introduction}
\label{sec:intro}
Bloch oscillations, Landau--Zener (LZ) tunneling, and Wannier--Stark ladders~\cite{Landau,Zener,Stueckelberg,Majorana32,Leo92,Dahan-Peik,Wilkinson,Anderson98,Glueck02}, are fundamental quantum effects occurring in a system of electrons moving in a periodic potential and subjected to a constant electric field. Due to complications such as impurities, lattice vibrations, and multiparticle interactions, clean observations of these effects have been difficult~\cite{Leo03}. In recent years, ultra-cold atoms and Bose--Einstein condensates in optical lattices have been increasingly used to simulate solid state systems and the above mentioned phenomena~\cite{Dahan-Peik,Wilkinson,Anderson98,niu98,Morsch01,Roati04,Morsch06,Bloch08}. Optical lattices are nowadays easy to realize in the laboratory, and their parameters can be perfectly controlled both statically and dynamically, which makes them attractive model systems for crystal lattices. More complicated potentials can be realized by adding further lattice beams~\cite{Santos04,Clement06-Sanchez-Palencia08,Lugan07,Schulte01,Salger01,Salger09}. In fact, by superimposing laser beams from different directions and with slightly different wave-lengths, it is possible to generate many different three-dimensional lattice geometries~\cite{Morsch06,Bloch08}. The question arises of how to control the dynamics of particles by quasiperiodic potentials (possibly time-dependent or even stochastic ones).\\
\indent  In this paper, we present results on the Wannier--Stark system realized with ultracold atoms, forming a Bose--Einstein condensate, in an optical lattice~\cite{Dahan-Peik,Wilkinson,Anderson98,Morsch01,Cristiani02,JonaLasinio03,Sias07,Zenesini08,Zenesini09}. We compute the time dependence of the tunneling probability of the Bose--Einstein condensate atoms out of the ground band in which they were originally prepared. By changing the initial condition and the system parameters and introducing atom-atom interactions into the system, we are able to control the tunneling rate of the Bose--Einstein condensate to higher bands. Finally a controlled noise added to the system will be shown to be a further handle to engineer the interband tunneling.\\
\section{Landau--Zener tunneling in optical lattices}
\label{sec:1}
\indent \sloppy We study the temporal evolution of ultracold atoms loaded into a quasi one-dimensional optical lattice which can be a spatially periodic potential or a time-dependent stochastic potential, subjected to an additional static force, in the presence of weak atom-atom interactions~\cite{Zenesini09}. We use the following general form of three-dimensional Gross--Pitaevskii equation to model the temporal evolution of the atoms
\begin{equation}\begin{split}
\label{eqno1}
H&=-\frac{\hbar^{2}}{2M}\nabla^2_{\vec{r}}+W\left(\vec{r},t\right)+g_{3D}\left\vert\Psi(\vec{r},t)\right\vert^2+Fx;\\ 
&W\left(\vec{r},t\right)=V_{\rm trap}\left(\vec{r}\right)+V_{1}\left(x\right)+V_{2}\left(x,t\right),
\end{split}
\end{equation}
with lattice potentials along the longitudinal direction, $x$, as
\begin{subequations}
\begin{align}
V_{1}(x) &= \alpha V\sin^{2}\left(\frac{\pi x}{d_{\rm L}}\right); \label{eqno2a}\\
V_{2}\left(x,t\right) &=\alpha V\sin^{2}\left(\frac{\pi x}{d^{'}_{\rm L}}+\phi(t)\right).
\label{eqno2b}
\end{align}
\end{subequations}
\indent \sloppy $M$ is the mass of condensate atoms and $F$ is the Stark force. $V_{1}(x)$ and $V_{2}(x,t)$ are spatially periodic potentials with incommensurate lattice spacings $d_{\rm L}$ and $d^{'}_{\rm L}$, respectively. As will be shown below around Eq.~\eqref{eqno9}, the noise has a tendency to average over the second lattice, and therefore the amplitudes of the two lattices should be comparable. For convenience we chose equal amplitudes $\alpha V$ and $d_{\rm L}=426\,{\rm nm}$ and $d^{'}_{\rm L}=d_{\rm L}(\sqrt{5}-1)/2$ for the lattice constants. $V_{2}(x,t)$ is a time-dependent stochastic potential with a time-dependent stochastic phase $\phi(t)$, which we will characterize further down in section \ref{subsec:2}. The renormalization factor $\alpha$ is introduced to be able to compare the dynamics in the presence of the potential given by Eqs.~\eqref{eqno2a} and~\eqref{eqno2b} to the dynamics of the ``reference system" , i.e. the dynamics in the potential $W\left(\vec{r}\right)=V_{\rm trap}\left(\vec{r}\right)+V\sin^{2}\left(\frac{\pi x}{d_{\rm L}}\right)$. $\alpha$ will be chosen in such a way that the following standard deviations are equal: 
\begin{align}
&\left\langle\left(\sin^{2}\left(\frac{\pi x}{d_{\rm L}}\right)-\left\langle \sin^{2}\left(\frac{\pi x}{d_{\rm L}}\right)\right\rangle_{x}\right)^2\right\rangle_{x}\nonumber\\  &= \left\langle\left( V_{\rm eff}\left(x\right)-\left\langle V_{\rm eff}\left(x\right)\right\rangle_{x}\right)^2\right\rangle_{x}\,,
\label{eqno3}
\end{align} 
where the effective potential $V_{\rm eff}(x)$ will be defined in section \ref{subsec:2}. The average $\langle\cdot\rangle$ is an integral over space for a sufficiently large $L$, i.e., $\langle\cdot\rangle=1/L\int_{-L/2}^{L/2}\cdot\,dx$. The third term in the Hamiltonian is the nonlinearity, which makes the equation different from the Schr\"odinger equation. $\Psi(\vec{r},t)$ represents the condensate wave function and $\left\vert\Psi(\vec{r},t)\right\vert^2$ the local atomic density. $g_{\rm 3D}=\frac{4\pi\hbar^2 a_{\rm s}N_{0}}{M}$ is the coupling constant which is proportional to the scattering length $a_{s}$ and determines the strength of atom-atom interactions, where $\frac{4\pi\hbar^2 a_{\rm s}}{M\,E_{R}}\approx2.45\times 10^{-21}\;{\rm m^3}$, with $a_s= 53\times 10^{-10}\;{\rm m}$ and $M=1.44\times 10^{-25}\;{\rm kg}$ for rubidium 87. $N_{0}$ is the number of atoms in the condensate. The recoil energy $E_{\rm{R}}=p^{2}_{\rm{R}}/2M$ and the recoil momentum $p_{\rm{R}}=\pi\hbar/d_{\rm L}$ are the characteristic energy and momentum scales for our system. Moreover, we define $V_{0}=V/E_{\rm R}$ and $F_{\rm 0}=Fd_{\rm L}/E_{\rm R}$ as dimensionless quantities in this energy unit. Experimentally, the initial state is the relaxed condensate wave function prepared in the confining potential given by a harmonic trap $V_{\rm trap}=\frac{1}{2}m\left(\omega_{\rm\rho}^2\rho^2+\omega_{\rm x}^2 x^2\right)$, with $\omega_{\rm x}\ll\omega_{\rm\rho}$ for a quasi 1D situation, and then loaded adiabatically into the optical lattice when the Stark force $F$ equals zero. Then  $\omega_{\rm x}$ is either switched off or relaxed to a small value $\omega_{\rm x,rel}$ and the Stark force $F$ is simultaneously switched on to induce the dynamics. In the following sections we study the system in the presence/absence of different terms of the above Hamiltonian.\\
    \begin{figure}
    \begin{center}
   \includegraphics[width=1.0\linewidth,angle=0]{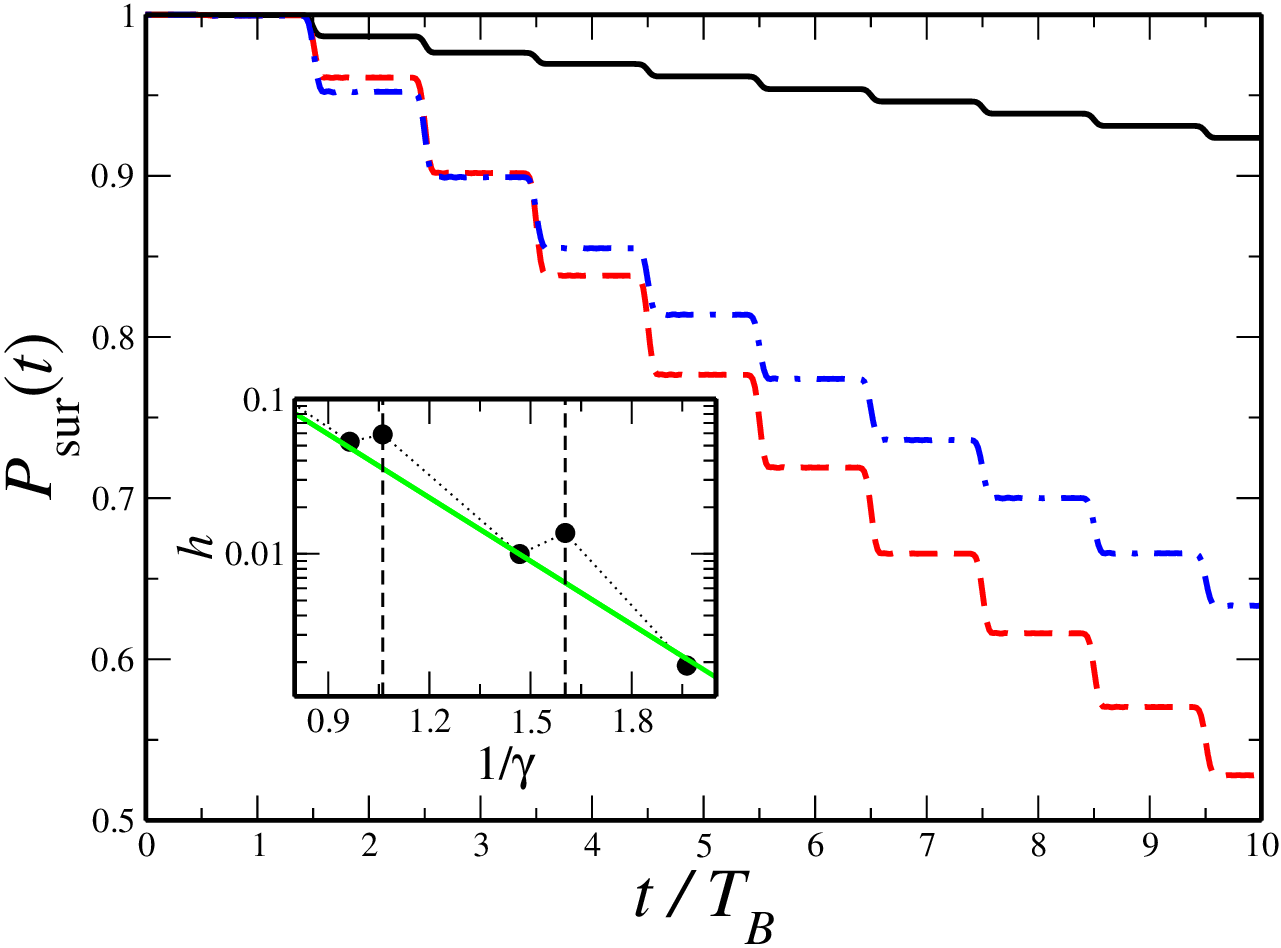} 
   \caption{\small Time resolved survival probability of the BEC in the ground band, for a lattice depth $V_{0}=4\;E_{\rm R}$, and a narrow width of the initial momentum distribution of the BEC cloud, $\Delta p/2p_{\rm R}=0.05$ (trap frequency $\omega_{\rm x}=2\pi\times 50\;\rm Hz$), and various Stark forces. For non-RET condition: $F_{0}=1.07$ (solid line) and $F_{0}=1.63$ (dash-dotted line); for RET condition: $F_{0}=1.48$ (dashed line). Inset: height of one step (of data as shown in the main panel) for fixed $V_{0}=4\;E_{\rm R}$ and various $\gamma$. The step height as predicted by Eq.~\eqref{eqno4}(solid line) in comparison with the numerically obtained values (filled circles). A significant deviation from the LZ prediction can be observed at values of $\frac{1}{\gamma}$ corresponding to RET conditions and marked by the vertical dashed lines. }
    \label{fig:1}  
    \end{center}
    \end{figure}
\subsection{Noise free case} 
\label{subsec:1}
\subsubsection{Linear case -- Wannier--Stark problem}
\label{subsubsec:1}
\begin{figure}
   \begin{center}
   \includegraphics[width=\linewidth,angle=0]{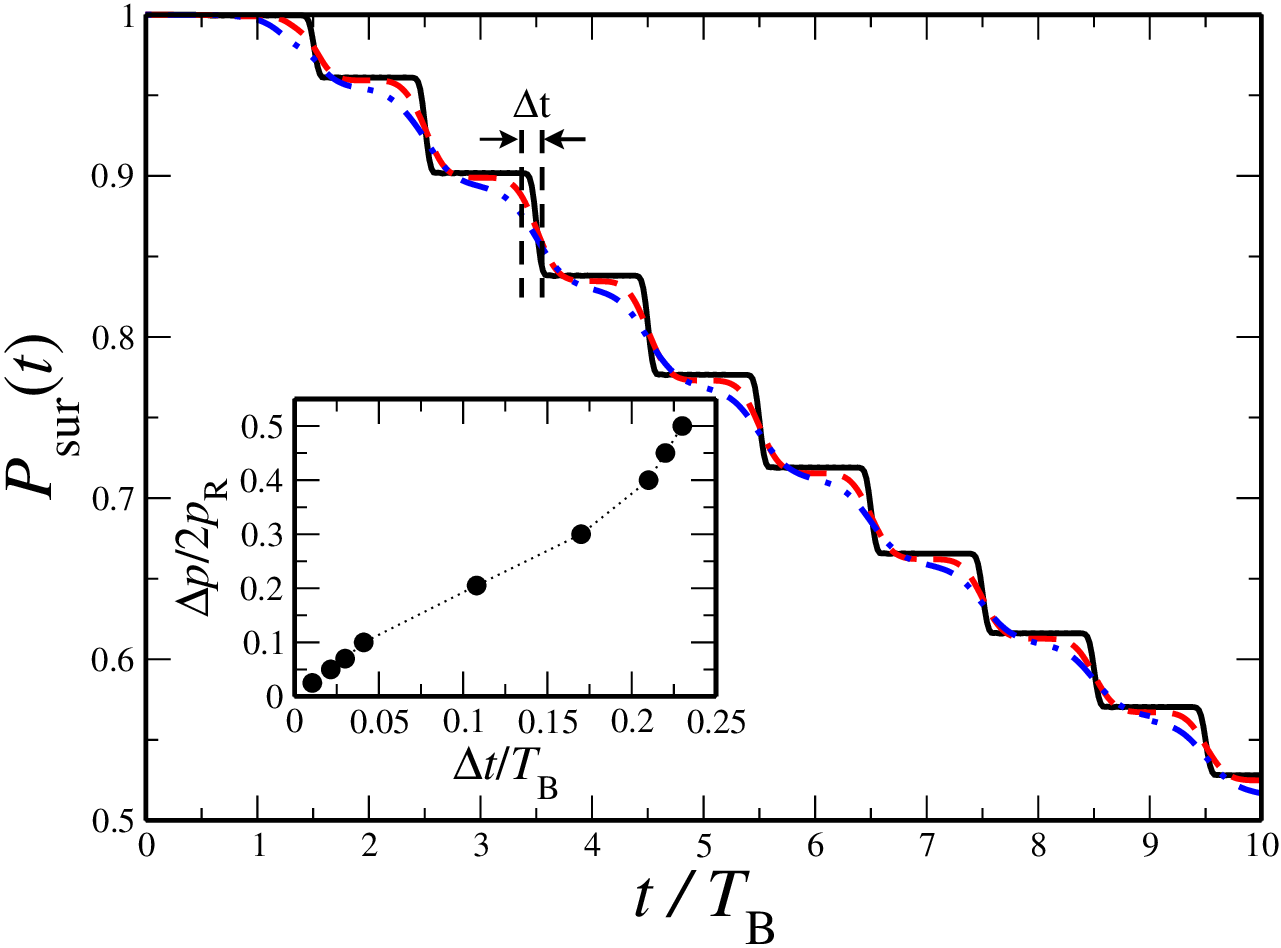}
   \caption{ \small The survival probability of the BEC in the ground band for a lattice depth $V_{0}=4.0\;E_{\rm R}$, a constant Stark force $F_{0}=1.48$, and various widths of the initial momentum distribution of the BEC cloud, $\Delta p/2p_{\rm R}=0.05$ (solid line); $\Delta p/2p_{\rm R}=0.2$ (dashed line); $\Delta p/2p_{\rm R}=0.3$ (dash-dotted line). Inset: step width $\Delta t$ for several $\Delta p$ (filled circles). The step width $\Delta t$ is the distance between the two vertical dashed lines shown in the main panel for the case of $\Delta p/2p_{\rm R}=0.05$. }  
   \label{fig:2} 
   \end{center}
   \end{figure}

\indent \sloppy For zero or small $\omega_{\rm x,rel}$ (c.f., e.g.,~\cite{Wimberger05}) and negligible atom-atom interactions, when $V_{2}(x,t)=0$ and $\alpha=1$, the above Hamiltonian in Eq.~\eqref{eqno1} describes the dynamics of atoms in a tilted periodic potential which is the well-known single-particle Wannier--Stark problem. Without the nonlinearity term, we can use the 1D version of Eq.~\eqref{eqno1} for our simulations. In the presence of $F$, the quasimomentum of a condensate (initially prepared at the center of the Brillouin zone in the ground band) scans the lower band in an oscillating motion periodically -- so-called Bloch oscillations -- with the Bloch period $T_{\rm B}=2\hbar (F d_{\rm L})^{-1}$. At the edge of the Brillouin zone, where the gap between the ground and the first excited band $\Delta E$ of the $F=0$ system (increasing with $V_{0}$~\cite{Holthaus00}), acquires its minimum value, a tunneling of the condensate to the first excited energy band may occur. The tunnelled atoms escape from the system through successive tunneling events across the much smaller band gaps between the upper bands. This phenomenon is known as the LZ tunneling. LZ theory predicts a decay rate 
\begin{equation}
 P_{\rm LZ}=e^{-\frac{\pi}{\gamma}},
\label{eqno4}
\end{equation}
where $\gamma$ is the adiabaticity parameter and $\gamma\approx\frac{32 F_{0}}{\hbar V_{0}^2}$~\cite{Holthaus00}. In order to study the LZ prediction for our system we need to access the decay rate of the population from the ground band. In that respect, we compute the time-dependence of the probability of the condensate to remain in the ground band in which it has been initially prepared. Such a survival probability is best measured in momentum space. From the time-dependent momentum distribution we can determine $P_{\rm sur}(t)$ by projection of the evolved state $\tilde{\Psi}(\vec{p},t)$ on to the support of the initial state~\cite{Wimberger05,Tayebirad10}
\begin{equation}
P_{\rm sur}(t)=\int_{-\infty}^{\infty}\int_{-\infty}^{\infty}dp_y\,dp_z\int_{-p_{\rm c}}^{\infty} dp_x \left|\Psi(\vec{p},t)\right|^{2},
\label{eqno5}
\end{equation}
where $p_{\rm c}$ is an ad hoc cut-off momentum. In our calculation we chose $p_{\rm c}= 3 p_{\rm rec}$. Then Eq.~\eqref{eqno5} starts to measure the wave packet leaving the ground-band with an artificial delay of one Bloch period, since by the acceleration theorem the average momentum  is proportional to time (more details can be found in~\cite{Wimberger05,Tayebirad10}). A very nice step structure -- local deviation from the overall exponential decay -- can be seen in the survival probability (see Figures~\ref{fig:1} and \ref{fig:2}). Such step structures reflect the above mentioned phenomena, i.e., Bloch oscillations and LZ tunneling. One can see that the tunneling events from the ground band to the next band occur after each Bloch period when the wave packet is at the edge of the Brillouin zone. It is possible to control the LZ tunneling of the Bose--Einstein condensate from the ground band in the linear system and in the absence of atom-atom interactions by
\begin{itemize}
 \item changing the system parameters, such as external force $F_{0}$, and the amplitude of the optical lattice $V_{0}$ (see Figure~\ref{fig:1}),
 \item exploiting resonantly enhanced tunneling (RET) between degenerate Wannier--Stark states at $\Delta E \approx n\times Fd_{\rm L}$ with $n$ being an integer number (see Figure~\ref{fig:1})~\cite{Glueck02},
 \item changing the initial condition by changing the trap frequencies and hence preparing the Bose--Einstein condensate with different widths $\Delta p$ of its initial momentum distribution (see Figure~\ref{fig:2}).
\end{itemize}
\indent \sloppy As seen in Figure~\ref{fig:1}, increasing the tilting force leads to more and more tunneling of the atoms from the ground band. On the other hand, depending on the system parameters, one can tune into a special condition for which the rate of the tunneling is enhanced. This occurs when an integer multiple of the energy scale of the tilting force $Fd_{\rm L}$ matches the energy difference between the initial state and the final state, i.e., approximately the band gap $\Delta E$. This phenomenon has been observed experimentally~\cite{Sias07,Zenesini08} and is called resonantly enhanced tunneling (RET). A deviation from the LZ prediction is expected in this case. In order to see whether the tunneling probability given by the standard LZ tunneling probability correctly predicts the height of a step corresponding to a single tunneling event, we fit a step function to our step-like survival probability and extract the height of each step. The result and the comparison to the LZ prediction is shown in the inset of Figure~\ref{fig:1}. When the system parameters are in the RET condition (e.g., $1/\gamma\approx 1.05$ and $1.6$), the height of the steps of the survival probability increases and shows a significant deviation from the LZ prediction given in Eq.~\eqref{eqno4}. This behavior is seen in Figure~\ref{fig:1} where the survival probability for $F_{0}=1.48$ decays much faster than for the other two cases. Since the derivation of Eq.~\eqref{eqno4}~\cite{Holthaus00} does not take into account the actual Wannier--Stark level structure, which is necessary to describe the RET condition, Eq.~\eqref{eqno4} or the survival probability derived from it cannot describe the enhancement of the tunneling probability due to RET.\\
\indent \sloppy The other parameter which can affect the survival probability is the width $\Delta p$ of the initial momentum distribution. By changing the trap frequencies $\omega_{x}$ and $\omega_{\rho}$ one can prepare the initial distribution with different widths. Figure~\ref{fig:2} demonstrates the dependence of the width of the steps $\Delta t$ on $\Delta p$. The steps are smooth and partly washed out since the wave packet reaches the edge of the Brillouin zone earlier when it has a broader initial momentum distribution. Still the remnants of the steps cause a local deviation from the exponential decay of the tunneling probability. Nevertheless, the survival probability exhibits an exponential decay globally in time, i.e., on large time scales, for all the mentioned cases.\\
\subsubsection{Non-linear case -- Gross--Pitaevskii equation}
\label{subsubsec:2}
\begin{figure}
    \begin{center}
    \includegraphics[width=\linewidth,angle=0]{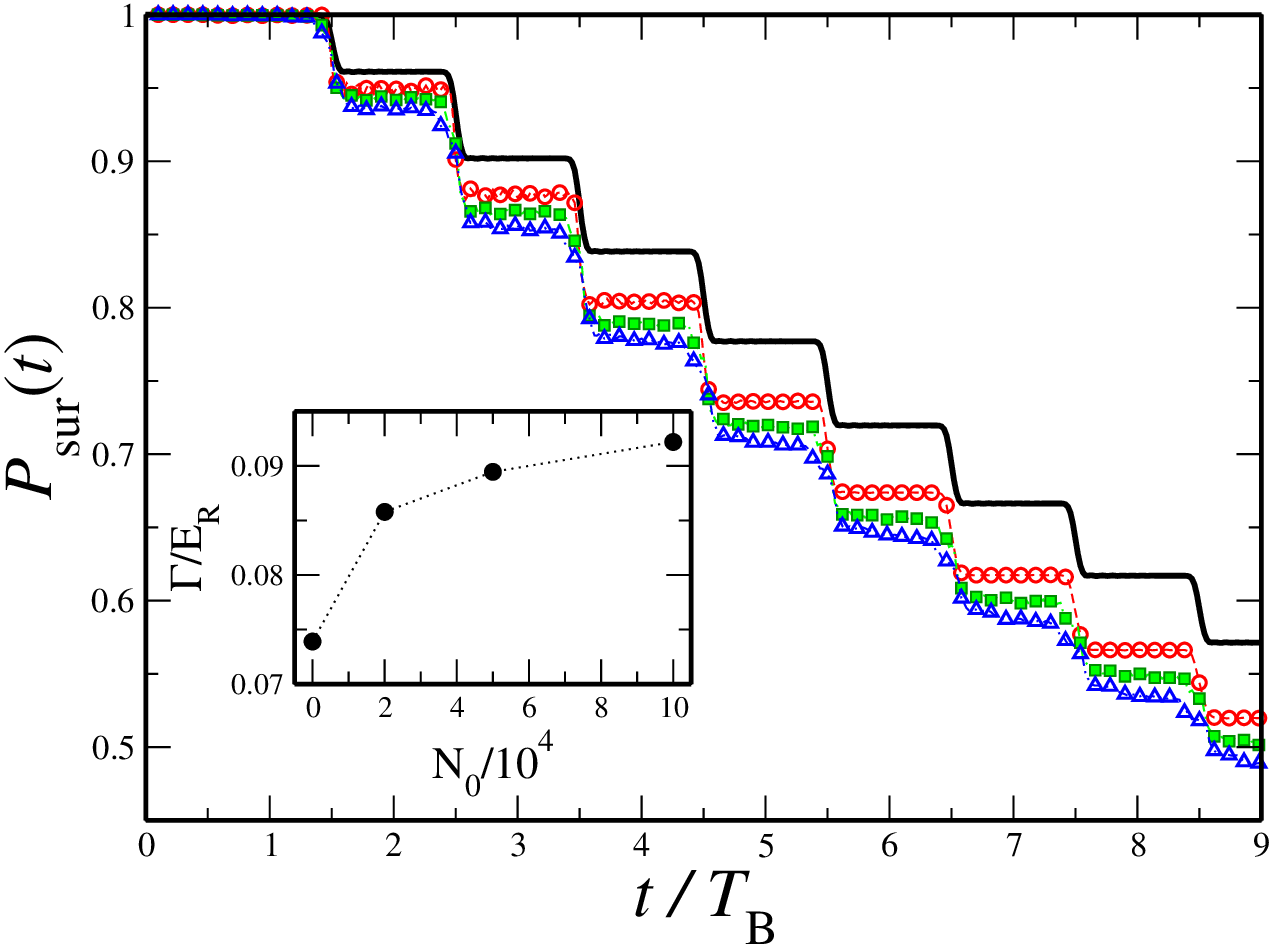}
    \caption{ \small Time resolved survival probability in the ground band for the nonlinear case with $V_{0}=4.0\;E_{\rm R}$, $F_{0}=1.48$ (RET), a narrow width of the initial momentum distribution of the BEC cloud, $\Delta p/2p_{\rm R}=0.05$, and composed of the following number of atoms prepared in a trap with $\omega_{\rm x}=2\pi\times 50\;\rm Hz$ and $\omega_{\rm \rho}=2\pi\times 100\;\rm Hz$: $N_{0}=2\times10^4$, $C\approx0.2$ (open circles); $N_{0}=5\times10^4$, $C\approx0.32$ (filled squares); $N_{0}=1\times10^5$, $C\approx0.42$ (open triangles) as compared to the linear case $g_{\rm 3D}=0$, (solid line). Inset: Decay rate of the survival probability $\Gamma$ at RET condition vs. $N_{0}$ (filled circles).}
    \label{fig:3} 
    \end{center}
    \end{figure}

\indent \sloppy In the regime of weak atom-atom interactions the effect of interactions is studied in the mean-field regime based on the Gross--Pitaevskii equation. The 3D Gross--Pitaevskii equation can describe the dynamics of the entire Bose--Einstein condensate in terms of an equation of motion for a single particle wave function. Therefore, the following nonlinear equation describes the dynamics of interacting Bose--Einstein condensate atoms in a tilted periodic potential ($V_{2}(x,t)=0$) with $\alpha=1$:
\begin{equation}
\label{eqno6}
H=-\frac{\hbar^{2}}{2M}\nabla^{2}_{\vec{r}}+V_{\rm trap}(\vec{r})+V_{1}\left(x\right)+g_{\rm 3D}\left\vert\Psi(\vec{r},t)^2\right\vert + Fx. 
\end{equation}
\indent \sloppy As mentioned in section~\ref{sec:1}, $g_{\rm 3D}$ is the coupling constant calculated from the s-wave scattering wavelength $a_{s}$ and the number of atoms in the condensate $N_{0}$. $\left\vert\Psi(\vec{r},t)^2\right\vert$ is the local atomic density. As an estimate for the nonlinear term in Eq.~\eqref{eqno6}, we define $C\equiv\frac{g_{\rm 3D}\left\vert\Psi(\vec{r},t)^2\right\vert_{\rm peak}}{E_R}$ at the peak density of the initial state. The following effects can be seen by increasing the strength of a repulsive nonlinearity ($g_{\rm 3D}>0$) in the system:
\begin{itemize}
\item enhancement of the tunneling rate,
\item deviation from mono-exponential decay,
\item washed out steps (corresponding to damped Bloch oscillations).
\end{itemize}
\indent\sloppy According to our results shown in Figure~\ref{fig:3} for a RET case and the experimental results in~\cite{Sias07,Zenesini08}, the temporal behavior of the survival probability depends on the strength of atom-atom interactions. As can be found with more detail in~\cite{Sias07,Wimberger05}, the enhancement of decay rate is generic for repulsive interactions. The scaling of the decay rate as a function of nonlinearity is yet more interesting in the RET case (see the inset of Figure~\ref{fig:3}). We can quantify the decay rate $\Gamma$ of the survival probability by globally fitting an exponential decay function to the step-like curves of the survival probability. Such rates, for various nonlinearities, are depicted in the inset of Figure~\ref{fig:3}. A repulsive interaction initially enhances the interband tunneling probability of the ultracold atoms~\cite{Wimberger05}. Since the tunneling events occurring at different integer multiples of the Bloch period are correlated by the presence of the nonlinearity, a clear deviation from the mono-exponential decay is observed making the definition of a global decay rate $\Gamma$ somewhat problematic~\cite{Schlagheck07}. A continuous change in the density of the condensate in time due to escaped particles from the system leads to a decreasing impact of the nonlinearity. Therefore, the time local rate of decay systematically decreases as the time increases. Additionally, the nonlinearity leads to a dephasing and damping of the Bloch oscillations, not discussed here, but a discussion of this phenomenon can be found in~\cite{Morsch01,Roati04,Morsch06,Gustavson08,Kolovsky}.\\
\subsection{Impact of a time-dependent stochastic potential (noisy Wannier--Stark problem)}
\label{subsec:2}
\begin{figure}
    \begin{center}
    \includegraphics[width=\linewidth,angle=0]{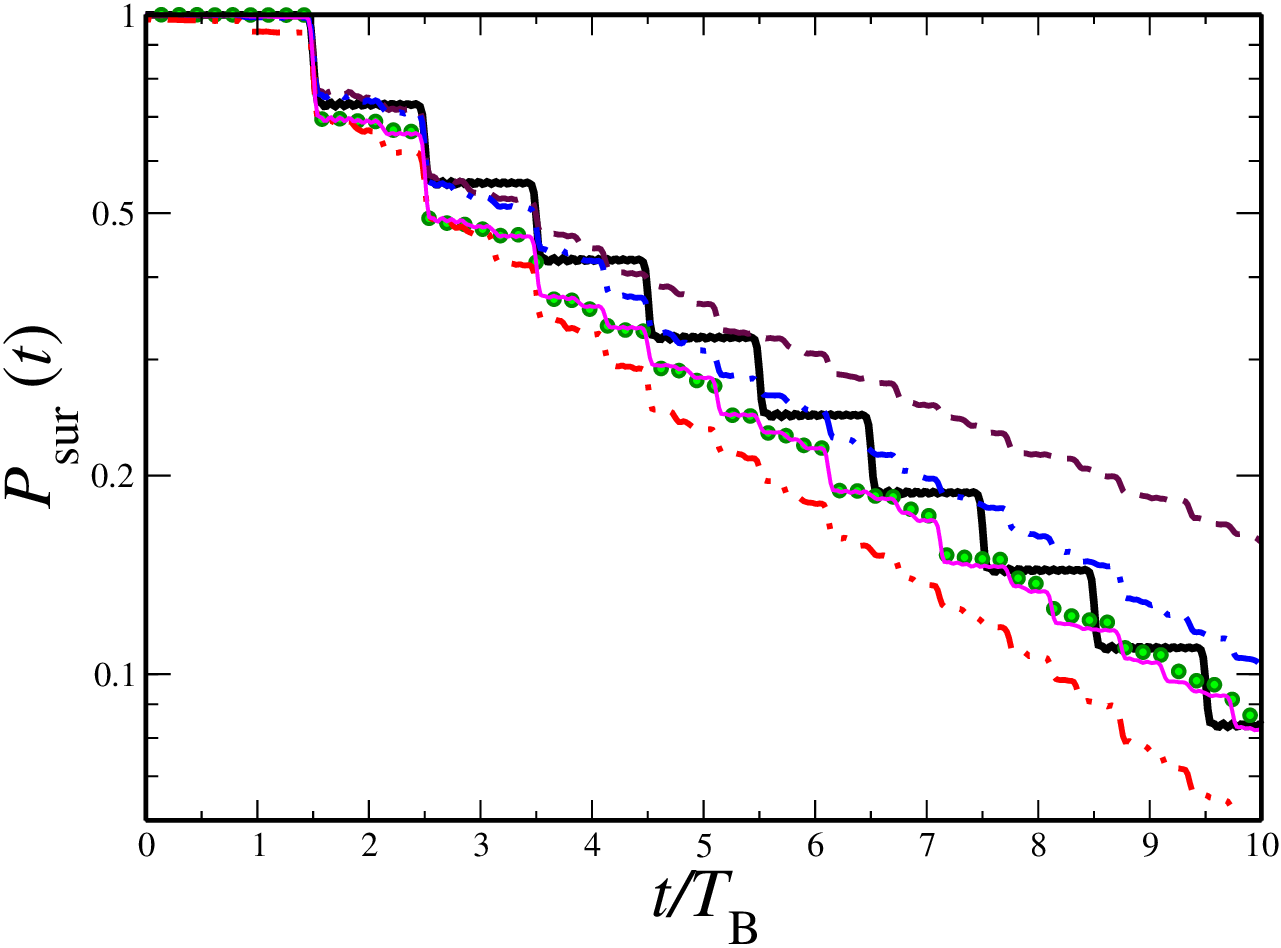}
    \caption{\small Time evolution of the survival probability for $V_{\rm 0}=2.5\;E_{\rm R}$, $F_{\rm 0}=1.5$, and $\alpha\approx0.82$; The thick solid line is the decay of the reference system defined by Eq.~\eqref{eqno1} with $g_{\rm 3D}=0$, $V_{2}(x,t)=0$ and $\alpha=1$; in the presence of exponentially correlated noise with $D/\tau\approx0.25$ and correlation times: $\tau=10\;T_{\rm B}$ (dashed line), $\tau=1\;T_{\rm B}$ (dot-dashed line), $\tau=0.05\;T_{\rm B}$ (dot-dot-dashed line), $\tau=0.001\;T_{\rm B}$ (thin solid line); results for the effective potential of Eq. \eqref{eqno9} are shown by the filled circles.}
    \label{fig:4} 
    \end{center}
    \end{figure}

\begin{figure}
    \begin{center}
    \includegraphics[width=\linewidth,angle=0]{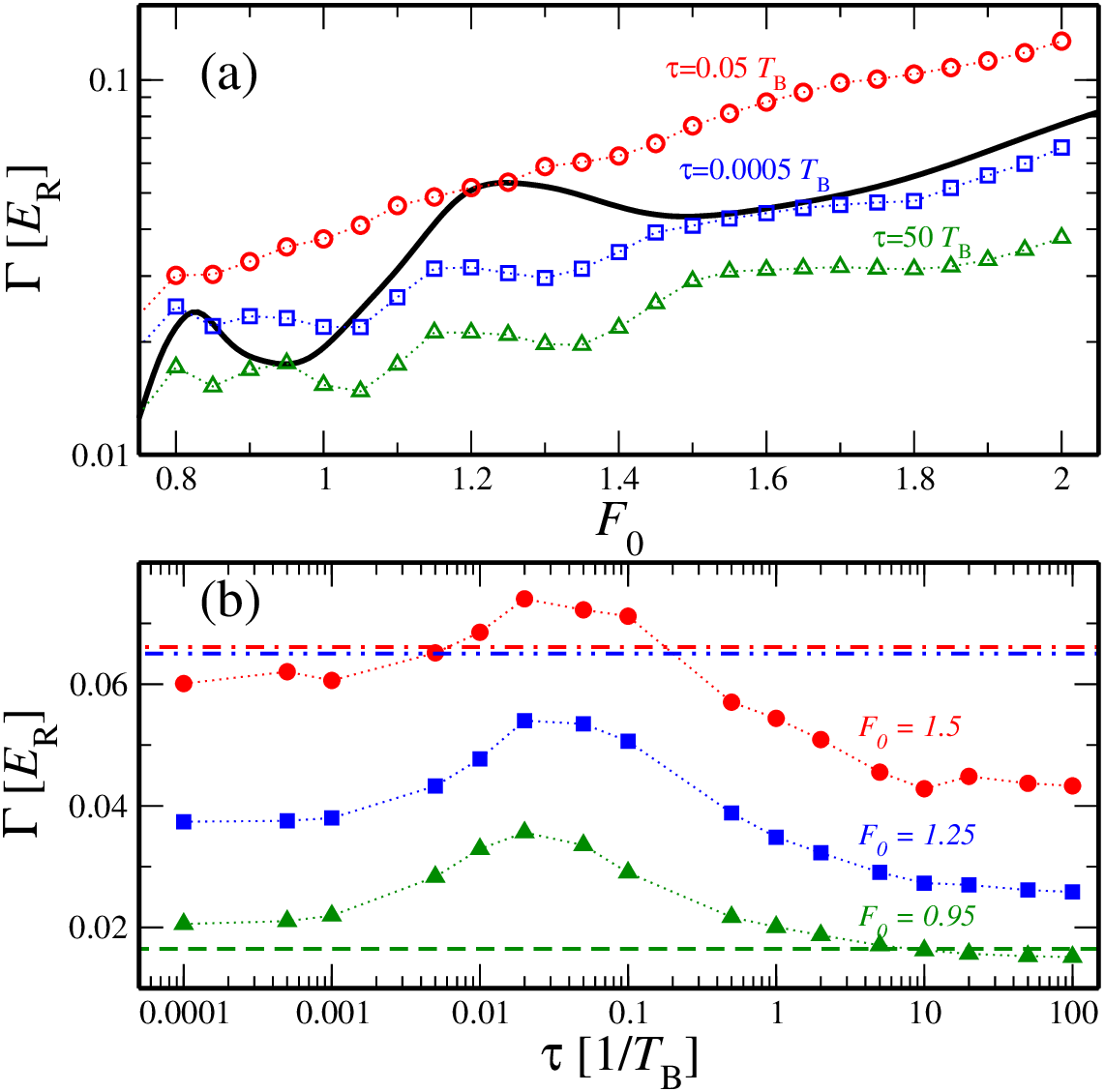}
    \caption{\small The decay rate of the survival probability for $V_{\rm 0}=2.5\;E_{\rm R}$, and $D/\tau\approx0.25$: (a) scan over $F_{\rm 0}$ for $\tau=0.0005\;T_{\rm B}$ (open squares), $\tau=0.05\;T_{\rm B}$ (open circles), $\tau=50\;T_{\rm B}$ (open triangles), and for the reference system (solid line); (b) scan over $\tau$ for $F_{\rm 0}=1.5$ (filled circles) and its reference system (dot-dot-dashed line), $F_{\rm 0}=1.25$ (filled squares) and its reference system (dot-dashed line), $F_{\rm 0}=0.95$ (filled triangles) and its reference system (dashed line).}
    \label{fig:5} 
    \end{center}
    \end{figure}

\indent \sloppy  Going back to the linear system, $g_{\rm 3D}=0$, the case of $V_{2}(x,t)\neq0$ leads to a time-dependent stochastic potential where we claim that we are able to control the dynamics of the Bose--Einstein condensate atoms by changing the characteristic parameters of the time-dependent stochastic phase $\phi(t)$. $\alpha$ is no longer 1 and we can calculate it as defined above by Eq.~\eqref{eqno3}, using the effective potential introduced below in Eq.~\eqref{eqno9}. We use correlated noise for the time-dependent phase $\phi(t)$ for the second lattice. A standard example is exponentially correlated noise, which is characterized by a single correlation time. Such a noise can be obtained from linearly filtered white noise as follows
   \begin{equation}
   \dot\phi=-\frac{\phi}{\tau}+\frac{\sqrt{2 D}}{\tau}\xi(t),
   \label{eqno7}
   \end{equation} 
where $\xi$ is a Gaussian white noise with zero mean and standard deviation equal to one. $\tau$ is the correlation time of the noise and the strength of the noise is given by the parameter $D$. The introduced noise has a zero mean ($\left<\phi(t)\right>=0$) and an exponential correlation function $\left<\phi(t)\phi(s)\right>=D/\tau\exp(- \left|t-s\right|/\tau)$. $\tau$ describes the time-scale of the fluctuations and $D/\tau$ its variance. The power spectrum of the exponentially correlated noise is given by a Lorentzian function as follows
   \begin{equation}
     S\left(\omega\right) = \frac{D}{\pi(1+\omega^{2}\tau^{2})}\;.  
     \label{eqno8} 
   \end{equation} 
\indent \sloppy Figure~\ref{fig:4} demonstrates the survival probability of the atoms in the ground band for various correlation times $\tau$ and a fixed value of the variance $D/\tau\approx0.25$ of the noise. The system parameters are $V_{\rm 0}=2.5\;E_{\rm R}$ and $F_{\rm 0}=1.5$ (non-RET condition). The solid line shows the time evolution of the survival probability for the reference system introduced by Eq.~\eqref{eqno1} with $g_{\rm 3D}=0$, $V_{2}(x,t)=0$ and $\alpha=1$. A nice step structure similar to the ones depicted in Figures~\ref{fig:1} and \ref{fig:2} is observed for this case. The other curves in Figure~\ref{fig:4} exhibit the time evolution of the survival probability for the temporally disordered system introduced in Eq.~\eqref{eqno1} with $g_{\rm 3D}=0$, $V_{2}(x,t)\neq 0$ and $\alpha\neq 1$. As seen for all the cases the step structure is washed out. Considering the characteristic parameters of the exponentially correlated noise, different regimes of the noise are introduced. The regime of slowly varying noise is recovered when the noise has a large correlation time compared with the time scales of the system (e.g., $T_{\rm B}$). The tunneling probability of the atoms in such a regime is suppressed (e.g., for $\tau=10\;T_{\rm B}$ shown by the dashed line in Figure~\ref{fig:4}). Decreasing the correlation time the rate of tunneling increases but, nevertheless, for $\tau\geq T_{\rm B}$ the noise suppresses the tunneling compared with the reference system (see the data for $\tau=1\;T_{\rm B}$ as shown by the dot-dashed line in Figure~\ref{fig:4}). For correlation times smaller than $T_{B}$, the noise recovers the regime of fast noise (e.g., for $\tau=0.05\;T_{\rm B}$) and causes an enhancement in the tunneling rate (dot-dot-dashed line) compared with the reference system (thick solid line). Surprisingly a further decrease of the correlation time decreases the tunneling rate of the atoms (e.g., for $\tau=0.001\;T_{\rm B}$ depicted by the thin solid line).\\ 
\indent \sloppy In order to understand the effect of the time-dependent stochastic potential on the system, we renormalize the potential. The time-dependent potential can be replaced by a suitable static effective potential in the limit of small $\tau$. Such an effective potential can be calculated integrating over all possible phases giving:
\begin{equation}
    V_{\rm eff}\left(x\right)=\alpha V\left[\sin^{2}\left(\pi x/d_{\rm L}\right)+\beta \sin^{2}\left(\pi x/d^{'}_{\rm L}\right)\right],
\label{eqno9}
   \end{equation}
with a renormalization factor for the second lattice $\beta=e^{-2D/\tau}$. Eq.~\eqref{eqno9} provides a time-independent potential which can be used to compute $\alpha$ as stated in Eq.~\eqref{eqno3}. As seen in Figure~\ref{fig:4} the survival probability for a small $\tau=0.001\;T_{\rm B}$ (thin solid line) shows perfect agreement with the results achieved using the effective potential of Eq.~\eqref{eqno9} (filled circles).\\
\indent \sloppy Keeping the noise parameters constant, we study the decay rate of the survival probability, $\Gamma$, of the condensate for various system parameters. The results of a scan over the Stark force are shown in Figure~\ref{fig:5}($a$). The simulations have been done for three values of the correlation time of the noise $\tau=0.0005, 0.05, 50\;T_{\rm B}$, keeping the variance of the noise constant (here $D/\tau\approx0.25$). The three correlation times are chosen from the left shoulder, the peak point and the right shoulder of curves in Figure~\ref{fig:5}($b$). The solid line depicts the decay rate of the reference system (given by Eq.\eqref{eqno1} with $g_{\rm 3D}=0$, $V_{2}(x,t)=0$ and $\alpha=1$). Noise in the system leads to a washing out of the RET peaks (present in the solid line of Figure~\ref{fig:5}($a$)) in the decay rate, and, dependent on the correlation time of the noise, a suppression (for very large and very small values of $\tau$) or an enhancement of the tunneling (e.g., for $\tau\approx 0.05\;T_{\rm B}$) can be obtained.\\
\indent \sloppy  We also ran a scan over the correlation time of the noise $\tau$, again keeping the variance of the noise constant ($D/\tau\approx0.25$). The decay rate of the survival probability for three amounts of the Stark force ($F_{0}=0.95, 1.25, 1.5$) is shown in Figure~\ref{fig:5}($b$). Comparing the symbols to the lines (which specify the decay rate in the corresponding reference system) one can realize that the decay rate is enhanced when the system parameters do \textit{not} fulfill the RET condition, i.e., in this case for $F_{0}=0.95$ and $1.5$. For the case of $F_{0}=1.25$ (RET) the tunneling rate is suppressed, and the symbols lie always below the reference line (dot-dot-dashed line). The enhancement is pronounced in the range of $\tau\approx0.005...0.2\;T_{\rm B}$, corresponding to an energy scale equal or larger to/than the band gap $\Delta E$. On the other hand, for small values of the correlation time, the noise recovers the regime of white noise, where the effective potential describes the dynamics of the system very well, and for large values of $\tau$, the system is in the regime of a slowly varying noise corresponding to energy scales that do not help to enhance the tunneling of atoms either.\\
\section{Conclusion}
\label{sec:3}
\indent \sloppy It is possible to control the Landau--Zener (LZ) tunneling probability of the ultracold atoms from the ground band in tilted optical lattices. This control is possible by changing the system parameters such as the lattice depth and the Stark force, or by changing the initial condition which is given by the initial width of the momentum distribution of the BEC. All the mentioned parameters can be easily tuned experimentally. Furthermore, our calculations showed that atom-atom interactions affect the LZ tunneling probability and a repulsive interaction typically leads to an enhancement of the LZ tunneling of the ultracold atoms from the ground band. Most interestingly, our results demonstrate that it is also possible to control the tunneling by adding noise to the system. By changing the noise parameters, the tunneling probability can be enhanced or suppressed. The noise can particularly enhance the tunneling probability when the system parameters are chosen far from the RET condition.\\

\begin{acknowledgement}
\indent \sloppy  We acknowledge funding by the Excellence Initiative by the German Research Foundation (DFG) through the Heidelberg Graduate School of Fundamental Physics (grant number GSC 129/1) and the Global Networks Mobility Measures. S. W. is grateful to the Heidelberg Academy of Sciences and Humanities for the Academy Award 2010 and to the Hengstberger Foundation for support by the Klaus-Georg and Sigrid Hengstberger Prize 2009. G. T. thanks the Landesgraduiertenf\"{o}rderung Baden-W\"{u}rttemberg for support.\\
\end{acknowledgement}

 \bibliographystyle{apsrmp}

\end{document}